\documentclass[epj,pacs]{svjour}
\usepackage{graphicx,amsmath,amssymb,bm}
\usepackage{color}
\usepackage[utf8]{inputenc}
\definecolor{purple}{rgb}{0.5,0,0.5}
\definecolor{blue}{rgb}{0.0,0,0.9}
\usepackage[colorlinks=true, pdfstartview=FitV, citecolor= purple, linkcolor = blue,urlcolor=blue]{hyperref}
\usepackage{amssymb}
\usepackage{graphics}
\usepackage{multirow}
\usepackage{epsfig}
\usepackage{rotating}
\usepackage{longtable} 
\usepackage{url}
\usepackage{pdflscape}
\usepackage{cancel}
\pdfoutput=1
\usepackage{csquotes}
\usepackage{placeins}
\usepackage{scalerel}
\usepackage{bbm}

\usepackage{array}
\usepackage{rotating}

\usepackage{comment}
\usepackage{widetext}
\usepackage{siunitx}
\usepackage{graphicx}
\begin{document}

\title{Linking Axions, the Flavor Problem and Neutrino Masses through a Flavored Peccei–Quinn Symmetry}

\author{Yithsbey Giraldo \thanks{yithsbey@gmail.com} \and Eduardo Rojas \thanks{eduro4000@gmail.com} and Juan C. Salazar \thanks{jusala@gmail.com} 
}    
\titlerunning{\emph{Linking Axions, Flavor, and Neutrinos}}
\authorrunning{Y. Giraldo, J. Salazar, E.Rojas.}
%
\offprints{}          
\institute{Departamento de F\'\i sica, Universidad de Nari\~no, A.A. 1175,  San Juan de Pasto, Colombia.}
\date{Received: date / Revised version: date}

\abstract{
Recent measurements by several experimental collaborations have reported deviations from Standard Model (SM) predictions in diphoton final states, potentially hinting at the existence of intermediate scalar resonances above the electroweak scale. Such anomalies can be naturally accommodated within SM extensions featuring an enlarged scalar sector. In particular, multi-Higgs doublet frameworks arise in Flavored Axion Models (FAMs), which have been proposed to explain the texture zeros of quark mass matrices. These models provide a unified description of quark masses and the Cabibbo–Kobayashi–Maskawa (CKM) mixing matrix while simultaneously addressing the strong CP problem.
In this work we study a concrete FAM realization augmented with Majorana masses for right-handed neutrinos, implementing a type-I seesaw mechanism.
In this model  the flavor structure is effectively determined by the vacuum expectation values of the scalar doublets and Yukawa couplings of order one. Within this framework, neutrino and axion mass scales are intrinsically connected, as the heavy right-handed neutrinos obtain their masses from the scalar field responsible for the spontaneous breaking of the Peccei–Quinn symmetry.
We further explore the phenomenological implications of the model, including constraints from flavor-changing neutral currents derived from semileptonic decays, as well as current experimental limits on the axion–photon coupling obtained from axion search experiments.
\PACS{{11.30.Hv}{Flavor symmetries}\and {12.60.-i}{models beyond the standard models}\and{14.80.Va}{Axions}\and {14.60.St}{Non-standard-model particles
Higgs bosons
neutrinos}\and {14.80.-j}{Other particles (including hypothetical)}}}
\maketitle


%

\maketitle

\section{Introduction}
The Peccei–Quinn (PQ) mechanism remains the most compelling resolution of the strong CP problem~\cite{Peccei:1977hh,Peccei:1977ur}. By promoting CP to an effectively exact symmetry of QCD at low energies, the spontaneous breaking of a global $U(1)_{\rm PQ}$ gives rise to a pseudo–Nambu–Goldstone boson—the axion—whose dynamics relax the QCD vacuum angle to zero~\cite{Weinberg:1978ma,Wilczek:1978ry}. Concrete ultraviolet realizations such as the KSVZ and DFSZ frameworks embed the axion into extensions of the Standard Model (SM) with additional scalar fields and fermion charges~\cite{Kim:1979if,Shifman:1979if,Zhitnitsky:1980tq,Dine:1981rt}. Comprehensive reviews of axion theory and phenomenology can be found in Refs.~\cite{Kim:2008hd,DiLuzio:2020wdo}.

Beyond solving strong CP, the PQ idea interfaces naturally with broader questions about the origin of mass scales in particle physics. Extended scalar sectors are a common ingredient in axion models (e.g. the DFSZ construction requires at least two Higgs doublets), opening paths to address the electroweak hierarchy and to generate textures in the Yukawa matrices that account for the observed fermion mass hierarchies and mixings. In particular, identifying (or aligning) flavor symmetries with $U(1)_{\rm PQ}$ can simultaneously control the axion couplings and reduce the number of free flavor parameters, yielding predictive textures for quarks and leptons~\cite{Wilczek:1982rv,Calibbi:2016hwq,Bjorkeroth:2018dzu,Bjorkeroth:2018ipq,DiLuzio:2017ogq,Rocha:2025ade}. Recent work continues to develop ``flavored'' PQ scenarios and $\nu$DFSZ variants that tie PQ breaking to the origin of neutrino masses and to controlled flavor structures~\cite{Ziegler:2019flavored,nuDFSZ:2025,Karan:2025pud}.

At the LHC, diboson final states are a sensitive probe of extended scalar sectors. A notable example is the rare SM decay $h\to Z\gamma$, where the ATLAS and CMS combination finds first evidence with a signal strength modestly above the SM expectation, albeit with current uncertainties~\cite{ATLASCMS:2023HtoZgamma}. Such loop–induced channels, together with precision measurements of $h\to\gamma\gamma$, constrain additional charged states and nonstandard scalar mixing. In parallel, direct searches for neutral resonances decaying to $\gamma\gamma$ or $Z\gamma$ have set stringent limits across a wide mass range~\cite{CMS:2024MergedDiphoton,ATLAS:2024LowMassDiphoton,ATLAS:2023ZaPhoton,CMS:2025B2GZH}. While no conclusive excess has emerged, several analyses report mild, localized deviations that motivate scalar explanations consistent with current bounds, especially in models predicting enhanced loop couplings to photons and $Z\gamma$.

These considerations motivate the class of models we study here: an extended scalar sector with several $SU(2)_L$ doublets and two singlets, organized by a flavored PQ symmetry. The presence of multiple Higgs doublets is not only natural in axion frameworks (as in DFSZ) but, when combined with appropriate PQ charge assignments, can generate realistic Yukawa textures through symmetry selection rules. 
The same doublets contain additional charged and neutral scalar states that can modify the loop–induced diboson rates, potentially accounting for hints in $Z\gamma$ and $\gamma\gamma$ while remaining compatible with existing collider limits. 
In general, constraints on the axion--photon coupling require the axion to be light, which in turn implies the presence of at least one scalar field acquiring a vacuum expectation value (VEV) of order $10^{6}\,\mathrm{GeV}$. Since several scalar couplings are unavoidable, this large VEV can raise the masses of the scalar states above the TeV scale, making the model incompatible with the Higgs-like resonances we aim to analyze, which typically appear at a few hundred~GeV. We find, however, that by appropriately choosing the trilinear terms in the scalar potential (dimension-three interactions involving only scalar fields), a phenomenologically viable scenario can be achieved.

Our analysis will delineate the parameter space where the axion solution to strong CP, flavor textures, and diboson phenomenology are simultaneously realized.

The manuscript is organized as follows. In Sec.~\ref{sec:textures} we review the quark and lepton mass–matrix textures employed throughout this work and express their real parameters in terms of the Standard Model fermion masses and two additional free parameters. In Sec.~\ref{sec:particle-content} we introduce the particle content of the model and specify the Peccei–Quinn charge assignments required to reproduce the mass textures discussed in Sec.~\ref{sec:textures}.

In Sec.~\ref{sec:yukawa} we determine the Yukawa couplings consistent with the observed charged–lepton and neutrino masses. Sec.~\ref{sec:scalars} is devoted to the scalar sector: we construct the most general scalar potential compatible with the symmetries of the model and perform a numerical scan of the parameter space to identify the regions yielding scalar–boson mass spectra consistent with the proposed diphoton excesses.

In Sec.~\ref{sec:axion-neutrino} we analyze the relation between the neutrino and axion mass scales. In Sec.~\ref{sec:constraints} we derive the relevant low–energy constraints. Finally, Sec.~\ref{sec:conclusions} summarizes our results and presents our conclusions.
%
 
\section{The Five texture-zero mass matrices\label{sec:textures}}
The motivation behind considering texture zeros in the Standard Model (SM) and its extensions is to reduce the number of independent parameters, thereby revealing potential correlations between masses and mixings in these models. The Yukawa Lagrangian provides the mechanism for fermion mass generation after spontaneous symmetry breaking. One initial simplification, without loss of generality, is to assume that the fermion mass matrices are Hermitian, which brings the number of free parameters in each quark and lepton sector down to 18. However, even with this reduction, there are still more parameters than needed to match the experimental data available in the literature. Since no comprehensive model exists to make accurate predictions, discrete symmetries are often employed to eliminate certain elements from the Yukawa matrix, resulting in texture zeros. In numerous studies, instead of introducing discrete symmetries, texture zeros are proposed as a more straightforward and practical alternative. This approach offers the advantage of allowing the selection of mass matrices in a way that optimizes the analytical treatment of the problem, while simultaneously enabling adjustments to the fermion masses and mixing angles.
\subsection{A Realistic Texture for Quark Mass Matrices}
It is important to note that six-zero textures in the Standard Model have already been ruled out, as their predictions fall outside the experimentally allowed ranges. However, five-zero textures for quark mass matrices remain a plausible option \cite{Verma:2017ppl,Xing:2019vks,Fritzsch:1999ee,Desai:2000bu,Ludl:2015lta,Ponce:2013nsa}. In particular, we focus on selecting five-zero textures that align well with the observed quark masses and mixing parameters, as demonstrated in previous works \cite{Giraldo:2011ya,Giraldo:2015cpp,Giraldo:2018mqi}.
{
\begin{equation}
\label{5.1y}
\begin{split}
M^{U}&=
\begin{pmatrix}
 0&0&C_u\\
0&A_u &B_u\\
C_u^*&B_u^*&D_u
\end{pmatrix},
\\
M^{D}&=
\begin{pmatrix}
 0&C_d&0\\
C_d^*&0&B_d\\
0&B_d^*&A_d
\end{pmatrix}.
\end{split}
\end{equation}}
The phases in \(M^{D}\) can be removed by applying a weak basis transformation (WBT)~\cite{Giraldo:2011ya,Branco:1988iq,Branco:1999nb}, which shifts these phases to the off-diagonal components of \(M^{U}\). This procedure enables the mass matrices in equation~\eqref{5.1y} to be expressed in an alternative form:
{
\begin{equation}
\label{eq2}
\begin{split}
M^{U}&
=
\begin{pmatrix}
 0&0&|C_u|e^{i\phi_{C_u}}\\
0&A_u &|B_u|e^{i\phi_{B_u}}\\
|C_u|e^{-i\phi_{C_u}}&|B_u|e^{-i\phi_{B_u}}&D_u
\end{pmatrix},
\\
M^{D}&=
\begin{pmatrix}
 0&|C_d|&0\\
|C_d|&0&|B_d|\\
0&|B_d|&A_d
\end{pmatrix},
\end{split}
\end{equation}}
By utilizing the trace and determinant of the mass matrices~\eqref{eq2}, both prior to and following the diagonalization procedure, the independent real parameters of \(M^{U}\) and \(M^{D}\) can be expressed in terms of their corresponding masses:
\begin{subequations}
\label{e3.4}
\begin{align}
\label{3.18}
 D_u&=m_u-m_c+m_t-A_u,\\
\label{34a}
|B_u|&=\sqrt{\frac{(A_u-m_u)(A_u+m_c)(m_t-A_u)}{A_u}},\\
\label{35a}
|C_u|&=\sqrt{\frac{m_u\,m_c\,m_t}{A_u}},
\\
 A_d&=m_d-m_s+m_b,\\
\label{34b}
|B_d|&=\sqrt{\frac{(m_b-m_s)(m_d+m_b)(m_s-m_d)}{m_d-m_s+m_b}},\\
\label{35b}
|C_d|&=\sqrt{\frac{m_d\,m_s\,m_b}{m_d-m_s+m_b}}.
\end{align}
\end{subequations}
One effective approach is to assume that the masses of the second-generation quarks are negative, corresponding to eigenvalues of \(-m_c\) and \(-m_s\). The parameter \(A_u\) remains free, and its value, which is set by the quark mass hierarchy, must fall within the specified range:
\begin{equation}
\label{eq5}
 m_u\le A_u\le m_t.
\end{equation}
The detailed step-by-step method for diagonalizing the mass matrices~\eqref{eq2} is outlined in Appendix~C reference~\cite{Giraldo:2020hwl}.
\section{PQ symmetry and the minimal particle content\label{sec:particle-content}}
\subsection{Yukawa Lagrangian and the PQ symmetry}
%

The texture-zeros of the mass matrices presented in equations \eqref{eq2} can be produced by applying a Peccei-Queen symmetry, $U(1)_{PQ}$, to the Lagrangian model as shown in Eq.\eqref{eq6}\cite{Bjorkeroth:2018ipq,Garnica:2019hvn,Ringwald:2015dsf}. As elaborated upon below, the minimal Lagrangian necessary to incorporate this symmetry is described in~\cite{Giraldo:2020hwl,Brivio:2017ije}.

\begin{align}
\mathcal{L}_{\text{LO}}& \supset 
(D_\mu\Phi^{\alpha})^\dagger D^\mu\Phi^{\alpha}
+\sum_{\psi}i\bar{\psi}\gamma^{\mu}D_\mu \psi
+\sum_{i=1}^{2} (D_\mu S_i)^\dagger D^\mu S_i\notag\\
&
- \Bigg(
\bar{q}_{Li}y^{D\alpha}_{ij}      \Phi^{\alpha}d_{Rj}  +
\bar{q}_{Li}y^{U\alpha}_{ij}\tilde\Phi^{\alpha}u_{Rj} \notag\\
&+
\bar{\ell}_{Li}y_{ij}^E  \Phi_3 e_{Rj}+
\bar{\ell}_{Li}y_{ij}^\nu\tilde\Phi_3\nu_{Rj} +\frac{1}{2}Y_{ij}^{N}\overline{\nu^c}_{\!\!Ri} \nu_{Rj}S_2^{\dagger} +\text{h.c} \Bigg)\notag\\
&+(\lambda_Q\bar{Q}_R Q_LS_2+\text{h.c})-V(\Phi,S_1,S_2)\,.
\label{eq6}
\end{align}
As demonstrated in reference~\cite{Giraldo:2020hwl}, the generation of quark mass textures necessitates the presence of at least four Higgs doublets, denoted as $\alpha=1,2,3,4$.

In the context of \eqref{eq6}, the indices $i$ and $j$ represent families (with an implied summation over any repeated indices). The superscripts $U$, $D$, $E$, and $N$ correspond to the up-type quarks, down-type quarks, electron-like fermions, and neutrino-like fermions, respectively. The covariant derivative in the Standard Model is given by $D_\mu = \partial_\mu + i\Gamma_\mu^{\text{SM}}$.

The charged leptons and neutrinos carry universal Peccei–Quinn charges and couple through Yukawa interactions to the scalar doublet $\Phi_3$. The right-handed neutrinos acquire Majorana masses, allowing the implementation of a Type-I seesaw mechanism. In an earlier work~\cite{Giraldo:2023osw}, we considered texture-zero structures for the lepton masses; however, this approach proved unnatural in the lepton sector. In the present model, we instead adopt the conventional Type-I seesaw solution to account for the smallness of neutrino masses.

For details on the scalar potential $V(\Phi,S_1,S_2)$, refer to Section~\ref{sec:scalars} and to the comprehensive information in reference~\cite{Giraldo:2020hwl}. In Eq.\eqref{eq6}, $\psi$ denotes the set of Standard Model fermion fields along with the heavy quark $Q$, which are identified in Tables \ref{tab:pcontent1} and \ref{tab:pcontent2}.

As illustrated in Table~\ref{tab:pcontent2}, the PQ charges for the heavy quark can be configured to permit interactions exclusively with the scalar singlet $S_2$. 

The $Q_{\text{PQ}}$ charges are assigned as follows: for left-handed quark doublets ($q_L$), we use $x_{q_i}$; for right-handed up-type quark singlets ($u_R$), $x_{u_i}$; for right-handed down-type quark singlets ($d_R$), $x_{d_i}$; for left-handed lepton doublets ($\ell_L$), $x_{\ell_i}$; for right-handed charged leptons ($e_R$), $x_{e_i}$; and for right-handed Dirac neutrinos ($\nu_R$), $x_{\nu_i}$ for each family ($i = 1, 2, 3$). A similar notation applies to the scalar doublets, $x_{\phi_{\alpha}}$ ($\alpha=1,2,3,4$), as well as the scalar singlets $x_{{S_{1,2}}}$.
\vspace{2cm}
\begin{widetext}    
\begin{table}[h]
\begin{center}

\scalebox{0.9}{\begin{tabular}{|ccccc|c|c|c|c|}
\hline  
Particles & Spin &$SU(3)_C$ &$SU(2)_L$ &   $U(1)_Y$  &$U_{\text{PQ}}(i=1)$& $U_{\text{PQ}}(i=2)$& $U_{\text{PQ}}(i=3)$&    $Q_{\text{PQ}}$\\
\hline
$q_{Li}$  & 1/2  & 3        &  2  & 1/6 &  $-2 s_1 + 2 s_2 + \alpha_q$  & $-s_1 + s_2 + \alpha_q$   & $\alpha_q$   & $x_{q_i}$     \\  
$u_{Ri}$  & 1/2  & 3        &  1  & 2/3  &  $s_1 + \alpha_q$ &  $s_2 + \alpha_q$  &  $-s_1 + 2 s_2 + \alpha_q$  & $x_{u_i}$     \\
$d_{Ri}$  & 1/2  & 3        &  1  & $-1/3$ &  $2 s_1 - 3 s_2 + \alpha_q$ &  $s_1 - 2 s_2 + \alpha_q$  & $-s_2 + \alpha_q$   & $x_{d_i}$     \\
$\ell_{Li}$  & 1/2  & 1        &  2  &   $-1/2$    & $\frac{x_{Q_R}-x_{Q_L}}{2}+s_1-2s_2$    & $\frac{x_{Q_R}-x_{Q_L}}{2}+s_1-2s_2$    &$\frac{x_{Q_R}-x_{Q_L}}{2}+s_1-2s_2$   & $x_{\ell_i}$    \\  
$e_{Ri}$  & 1/2  & 1        &  1  &    $-1$   & $\frac{x_{Q_R}-x_{Q_L}}{2}+2s_1-4s_2$    & $\frac{x_{Q_R}-x_{Q_L}}{2}+2s_1-4s_2$   & $\frac{x_{Q_R}-x_{Q_L}}{2}+2s_1-4s_2$    & $x_{e_i}$    \\
$\nu_{Ri}$& 1/2  & 1        &  1  &  0     & $\frac{x_{Q_R}-x_{Q_L}}{2}$   & $\frac{x_{Q_R}-x_{Q_L}}{2}$    &  $\frac{x_{Q_R}-x_{Q_L}}{2}$  & $x_{\nu_i}$   \\
\hline
\end{tabular}}
\caption{Particle content. The subindex  $i=1,2,3$ stand for the family number in the interaction basis.  Columns 6-8 are the Peccei-Quinn charges, $Q_{PQ}$, for each family of quarks and leptons in the SM. $s_1, s_2$ and $\alpha$ are real parameters, with $s_1\ne s_2$.}
\label{tab:pcontent1}
\end{center}
\end{table}
\begin{table}
\begin{center}
\begin{tabular}{|ccccccc|}
\hline   
Particles & Spin &$SU(3)_C$ &$SU(2)_L$ &$U(1)_Y$&     $U_{\text{PQ}}$              &$Q_{\text{PQ}}$\\
\hline
$\Phi_{1}$  & 0  & 1        &  2       &   1/2  &     $s_1$         &$x_{\phi_1}$   \\  
$\Phi_{2}$  & 0  & 1        &  2       &   1/2  &      $s_2$        &$x_{\phi_2}$   \\  
$\Phi_{3}$  & 0  & 1        &  2       &   1/2  &       $-s_1 + 2 s_2$             &$x_{\phi_3}$   \\
$\Phi_{4}$  & 0  & 1        &  2       &   1/2  &      $-3 s_1 + 4 s_2$            &$x_{\phi_4}$   \\  
$Q_L$       & 1/2& 3        &  1       &    0   &      $x_{Q_L}$    &$x_{Q_L}$      \\ 
$Q_R$       & 1/2& 3        &  1       &    0   &     $x_{Q_R}$     &$x_{Q_R} $     \\  
$S_1$       & 0  & 1        &  1       &    0   &      $s_1-s_2$    &$x_{_{s_1}}$   \\ 
$S_2$       & 0  & 1        &  1       &    0   &   $x_{Q_R}-x_{Q_L}$& $x_{_{s_2}}$  \\ 
\hline
\end{tabular}
\caption{Beyond the SM fields and their respective PQ charges.
 The parameters $s_1, s_2$ are reals, with $s_1\ne s_2$  and $x_{Q_R}\ne x_{Q_L}$.}
\label{tab:pcontent2} 
\end{center}
\end{table}
\end{widetext}

In this study, the PQ charges attributed to both the quark and scalar sectors, along with the VEVs assigned to the scalar doublets, will mirror those assigned in\cite{Giraldo:2020hwl}, as detailed in Tables\eqref{tab:pcontent1} and~\eqref{tab:pcontent2}.

Our model incorporates two scalar singlets, $S_1$ and $S_2$, responsible for breaking the global symmetry $U(1)_{PQ}$.

The QCD anomaly associated with the PQ charges is described by:

\begin{align}
N = 2\sum_{i=1}^3 x_{q_i} - \sum_{i=1}^3 x_{u_i} - \sum_{i=1}^3 x_{d_i} + A_Q\ ,
\end{align}

where $A_Q = x_{QL} - x_{QR}$ accounts for the anomaly contribution from the heavy quark $Q$, which is an electroweak gauge group singlet. The left and right Peccei-Quinn charges are denoted by $x_{QL}$ and $x_{QR}$, respectively.

We express these charges as a function of $N$, ensuring that $N$ is non-zero, such that
\begin{align}\label{eq:parametrization}
s_1= \frac{N}{9}\hat{s}_1,\hspace{1cm}s_2=\frac{N}{9}\left(\epsilon+\hat{s}_1 \right),\ \  \text{with}
\hspace{0.5cm }\epsilon=1-\frac{A_{Q}}{N}\,, 
\end{align}
where $\hat{s}_1$ and $\epsilon$ are arbitrary real values.

To address the strong CP problem with $N \ne 0$ while also generating texture-zeros in the mass matrices, it is essential to ensure $\epsilon = \frac{9 (s_2 - s_1)}{N} \ne 0$.

In this context, concerning Flavor-Changing Neutral Currents (FCNC) observables, the parameters of interest are $\hat{s}_1$ and $\epsilon$.

This parameterization proves particularly useful, especially when the parameters $\alpha_q$ and $\alpha_{\ell}$ are not significant, because by setting $N$ and $f_a$, we can adjust $\hat{s}_1$ and $\epsilon$ while keeping $\Lambda{\text{PQ}} = f_a N$ constant. This allows the parameter space to be effectively reduced to two dimensions.
%
\section{Naturalness of Yukawa couplings\label{sec:yukawa}}
As demonstrated in Ref.~\cite{Giraldo:2020hwl}, the realization of five texture zeros in the quark mass matrices of Eq.~\eqref{5.1y}, enforced by a Peccei--Quinn (PQ) symmetry, requires the introduction of at least four scalar $SU(2)_L$ doublets. Once the electroweak symmetry is spontaneously broken, the up- and down-type quark mass matrices can be written as
\begin{align}\label{eq:yij}
M^{U}&=\hat{v}_{\alpha}\, y^{U\alpha}_{ij}= 
\begin{pmatrix}
       0 & 0 & y^{U1}_{13}\hat{v}_1\\[0.2cm]
       0 & y^{U1}_{22}\hat{v}_1 &  y^{U2}_{23}\hat{v}_2\\
       y^{U1^*}_{13}\hat{v}_1 & y^{U2^*}_{23}\hat{v}_2 & y^{U3}_{33}\hat{v}_3
\end{pmatrix},\notag\\
M^{D}&=\hat{v}_{\alpha}\, y^{D\alpha}_{ij}=
\begin{pmatrix}
0 & |y^{D4}_{12}|\hat{v}_4 & 0\\
|y^{D4}_{12}|\hat{v}_4 & 0 & |y^{D3}_{23}|\hat{v}_3\\
0 & |y^{D3}_{23}|\hat{v}_3 & y^{D2}_{33}\hat{v}_2    
\end{pmatrix},
\end{align}
where $\hat{v}_i \equiv v_i/\sqrt{2}$ are defined in terms of the vacuum expectation values of the scalar doublets.

It was further shown in Ref.~\cite{Giraldo:2020hwl} that the five-zero textures of Eq.~\eqref{eq2} allow most of the quark Yukawa couplings to be naturally of order unity, with the exception of $y^{U2}_{23}$, $y^{D3}_{23}$, and $y^{U1}_{13}$. Under this assumption, one obtains
\begin{align}
\label{eq28}
\hat{v}_1 &= 1.71~\text{GeV}, \qquad
\hat{v}_2 = 2.91~\text{GeV},\notag\\
\hat{v}_3 &= 174.085~\text{GeV}, \qquad
\hat{v}_4 = 13.3~\text{MeV}.
\end{align}
Although hermiticity is not automatically guaranteed in a generic multi-Higgs framework~\cite{CarcamoHernandez:2022vjk}, it is well motivated to impose it in the present context. First, in the Standard Model and in many of its extensions where right-handed fermions are $SU(2)_L$ singlets, the quark mass matrices can be chosen Hermitian without loss of generality. Second, this property remains valid even after implementing an additional PQ symmetry. Third, Hermitian textures permit the application of the weak-basis transformation (WBT) method~\cite{Giraldo:2011ya}. Finally, there exists an extensive body of phenomenologically viable Hermitian quark mass matrices in the literature. We emphasize that the mass matrices given in Eq.~\eqref{eq:yij} satisfy the Hermiticity condition.

\section{Scalar potential\label{sec:scalars}}

The scalar sector under consideration includes four Higgs doublets~($\Phi_{1,2,3,4}$) and two complex scalar singlets~($S_{1,2}$), embedded into the broader program of Higgs–like resonance searches at the LHC. The resulting spectrum—six CP--even, three CP--odd, and six charged scalars—offers a consistent interpretation of the \SI{95}{GeV} diphoton excess, the Standard-Model-like Higgs at \SI{125}{GeV}, and possible heavier scalar states extending up to the multi--TeV scale. Relevant decay modes, current collider constraints, and flavour observables have been analysed, with projections for LHC Run~3 and the HL--LHC also taken into account.

Motivated by multi--Higgs models with a global Peccei–Quinn~(PQ) symmetry aimed at solving the strong--CP problem~\cite{Peccei:1977hh}, the setup assumes an extended scalar sector preserving $SU(2)_L \times U(1)_Y \times U(1)_{\text{PQ}}$, where all doublets carry SM quantum numbers. The scalar fields are written as
\begin{align}
\label{eq:higgs1}
\Phi_\alpha =& 
\begin{pmatrix}
\phi_\alpha^{+}\\
\frac{v_\alpha+h_\alpha+i\eta_\alpha}{\sqrt{2}}
\end{pmatrix},
\hspace{1cm}
\tilde\Phi_\alpha=i\sigma_2 \Phi_\alpha^{*},
\hspace{0.5cm}\notag {\alpha=1,2,3,4,}\\
S_i=&\frac{v_{_{s_i}}+\xi_{s_i}+i\zeta_{s_i}}{\sqrt{2}};\hspace{1cm} i=1,2,
\end{align}
with vacuum expectation values satisfying the hierarchy $v_4\ll v_1, v_2\ll v_3\ll v_{s_2}$, as previously established in~\cite{Giraldo:2020hwl,Giraldo:2023osw}. These VEVs, along with suitable PQ charge assignments, induce Yukawa textures with predictive power in the quark sector, reproducing observed masses and CKM mixing while suppressing the number of free parameters. A vector--like quark, charged under PQ symmetry, acquires mass through the coupling
$
\mathcal{L}\supset \lambda_QS_2\bar Q_L Q_R +\text{h.c.},
$
which ensures the finiteness of QCD--PQ anomalies and helps control FCNC effects.

\begin{widetext}
\begin{figure}
\begin{center}
 \includegraphics[scale=0.5]{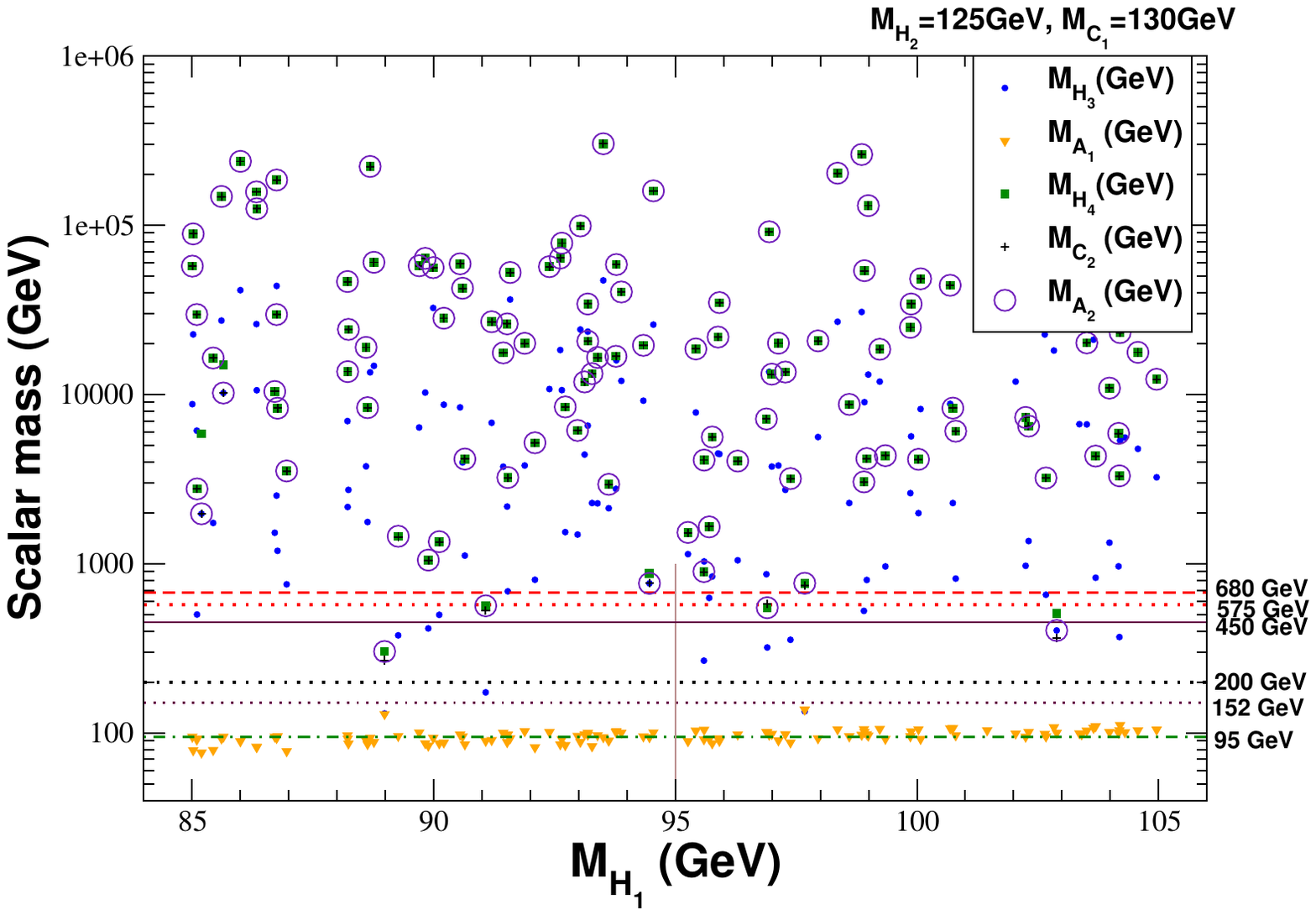}
\end{center}
\vspace{-1cm}
\caption{
To connect with current phenomenology we set
$
M_{H_1}=\SI{95}{GeV}$, $M_{H_2}=\SI{125}{GeV}$, $M_{C_1}=\SI{130}{GeV}$. Heavier scalars span the ranges
$M_{H_3}\in[\SI{200}{GeV},\SI{50}{TeV}]$, 
$M_{C_2}\in[\SI{300}{GeV},\SI{300}{TeV}]$, $M_{A_1}\in[\SI{95}{GeV},\SI{150}{GeV}]$.
Higher states are taken above the current LHC reach.}
\label{fig:escalares}	
\end{figure}
\end{widetext}

The singlet $S_2$ breaks the PQ symmetry at a high scale, $f_a^2=\sum_i(x_iv_i)^2\approx x_{s_2}^2v_{s_2}^2 \approx (10^{15})^2$~GeV$^2$ (see Appendix~\ref{sec:fa}). However, in contrast with earlier implementations~\cite{Giraldo:2020hwl,Giraldo:2023osw}, and despite the large value of $v_{s_2}$, it is possible to obtain a spectrum for scalar masses in the range of \SI{100}{GeV}, as suggested by the Higgs-like signatures observed at the LHC, as shown in the figure~\ref{fig:escalares}. A numerical scan over $v_{s_1}$ in the range \SI{50}{GeV}–\SI{40e3}{GeV} revealed that beyond this range, it becomes increasingly difficult to obtain scalar masses below \SI{1}{TeV}.
Due to the large value of $v_{s_2}$, the model predicts a light axion state. Since the axion decay constant scales as $f_a \propto v_{s_2}$, the axion mass behaves as $m_a =(7.5\times 10^{-2}\text{GeV})^2/f_a$, leading to a suppressed axion mass for sizable $v_{s_2}$.

The scalar potential compatible with the symmetries includes quadratic, quartic, and specific trilinear interactions:
{\footnotesize
\begin{eqnarray}\label{eq:scalar-potential}
V(\Phi_i,S_j) &=& \sum_{i=1}^4\mu_{i}^{2}\Phi_{i}^{\dagger}\Phi_{i} +\sum_{k=1}^2\mu_{s_k}^2 S_k^{*}S_k+\sum_{i=1}^4\lambda_{i}\left(\Phi_{i}^{\dagger}\Phi_{i}\right)^{2}\nonumber \\
&+&  \sum_{k=1}^2\lambda_{s_k}\left(S_k^{*} S_k \right)^{2}     + \sum_{i=1}^4\sum_{k=1}^2\lambda_{is_k}\left(\Phi_{i}^{\dagger}\Phi_{i}\right)\left(S_k^{*}S_k\right)\nonumber\\
&+& \sum_{\underbrace{i,j=1}_{i<j}}^4\bigg{(}\lambda_{ij}\left(\Phi_{i}^{\dagger}\Phi_{i}\right) \left(\Phi_{j}^{\dagger}\Phi_{j}\right)+J_{ij}\left(\Phi_{i}^{\dagger}\Phi_{j}\right) \left(\Phi_{j}^{\dagger}\Phi_{i}\right)\bigg{)}
\nonumber\\
&+& \lambda_{s_1 s_2}\left(S_1^{*}S_1\right)\left(S_2^{*}S_2\right)
+K_{1}\left(\left(\Phi_{1}^{\dagger}\Phi_{2}\right) \left(\Phi_{3}^{\dagger}\Phi_{2}\right) + h.c.\right)\nonumber\\
&+& K_{2}\left(\left(\Phi_{3}^{\dagger}\Phi_{4}\right) \left(\Phi_{3}^{\dagger}\Phi_{1}\right) + h.c.\right)\nonumber\\
&+&K_{3}\left(\left(\Phi_{3}^{\dagger}\Phi_{4}\right) S_1^2 + h.c.\right)+ K_{4}\left(\left(\Phi_{1}^{\dagger}\Phi_{3}\right) S_1^2 + h.c.\right)\nonumber\\
&+& F_1 \left( \left(\Phi_{2}^{\dagger}\Phi_{3}\right) S_1 +h.c.\right)+F_2 \left( \left(\Phi_{1}^{\dagger}\Phi_{2}\right) S_1 +h.c.\right)\nonumber\\
&+&\frac{1}{2} \left(m_{\zeta_{s_2}}\right)^2_{\text{SB}}\zeta^2_{s_2}+\frac{1}{2}\left(m_{\xi_{s_2}}\right)^2_{\text{SB}}\xi^2_{s_2}\ .
\end{eqnarray}}
Due to its PQ charge, $S_2$ does not couple trilinearly or quartically to the doublets, while $S_1$ does. The $F_i$ couplings arise from the PQ charge assignments and have mass dimension one. The singlet $S_2$ also provides mass to the heavy quark $Q$.

To preserve naturalness, the following dimensionless parameters in the potential~\eqref{eq:scalar-potential} were chosen in the range:
\begin{equation}
\lambda_{is_k},\lambda_{ij},J_{ij},\lambda_{s_1s_2},K_1, K_2, K_4 \in[-1.5,1.5]\ .
\end{equation}

Additionally, to ensure the scalar potential is stable—i.e., bounded from below—its stability is determined solely by the quartic terms in 
V:

\begin{equation}
\lambda_i,\lambda_{s_k} \in[0,1.5]\ .
\end{equation}

Despite being dimensionless, $K_3$ is subject to stronger constraints, leading to narrower intervals. Similar considerations apply to $F_1$ and $F_2$:
\begin{equation}
\begin{split}
F_1&\in[-45,10]\:\text{GeV}\ ,\\
F_2&\in[-600,600]\:\text{GeV}\ ,\\
K_3&\in[-1.5,0.2]\ .
\end{split}
\end{equation}

\noindent
{After electroweak symmetry breaking, one neutral CP--odd Goldstone $G^0$ and one charged Goldstone pair $G^\pm$ are absorbed by $Z$ and $W^\pm$, respectively. The remaining neutral CP--odd massless state is the axion $a$, the (pseudo-)Nambu--Goldstone boson associated with the spontaneous breaking of $U(1)_{\rm PQ}$. The physical spectrum then comprises six CP--even scalars    $H_{1\ldots6}$, four CP--odd states\\$\{A_{1},A_{2},A_{3},a\}$, and three charged scalar pairs $C^{\pm}_{1\ldots3}$.}\\

We define the real orthogonal matrix \(R\) that diagonalizes the CP-even mass matrix in the gauge basis
\[
\Phi=\big(h_1,h_2,h_3,h_4,\xi_{s_1},\xi_{s_2}\big):
R^{T} M_H^{2} R = \operatorname{diag}\!\big(m_{H_1}^2,\ldots,m_{H_6}^2\big).
\]
The entry \(R_{ik}\equiv \langle \Phi_i \mid H_k \rangle\) quantifies the admixture of \(\Phi_i\) in \(H_k\)\    .

For the hierarchy 
\(
v_{s_2}\gg v_{s_1}\gtrsim v_3\gg v_{1,2}\gg v_4,
\)
the states \(H_6\simeq S_2\) and \(H_5\simeq \Phi_4\) naturally decouple. 
The former does so because 
\(
m_{H_6}^2\simeq 2\,\lambda_{s_2}\,v_{s_2}^2,
\)
while the latter becomes heavy since 
\(
(M_H^2)_{44}\approx \frac{v_3}{2v_4}(K_{2}v_{1}v_{3}+K_{3}v_{s_1}^{2}),
\)
except for the unlikely case of a fine-tuned cancellation,
\(
K_{2}v_{1}v_{3}+K_{3}v_{s_1}^{2}\simeq 0,
\)
the entry \((M_H^{2})_{44}\) remains large. 
If such a cancellation were to occur, \((M_H^{2})_{44}\) would be suppressed and the mixing significantly enhanced; this corresponds to a tuned region of parameter space.
The mode \(H_2\) aligns with the SM Higgs boson,
\(
H_{2} \approx \Phi_{3},
\)
with tiny admixtures (corresponding to \(|R_{32}|\to 1\)), thereby reproducing the measured SM Higgs couplings. 
In contrast, the light state \(H_1\simeq 95~\mathrm{GeV}\) originates from the subspace spanned by \(h_1\) and \(h_2\), orthogonal to the direction associated with \(h_3\). 
This state can contain at most a small component
\(
|R_{31}| = |\langle \Phi_3| H_1 \rangle| \lesssim 0.243,
\)
which is sufficient to reproduce the \(\gamma\gamma\) signal strength without placing the SM Higgs couplings under tension.

The diphoton signature near \SI{95}{GeV} is mediated by the light CP-even state $H_1$. Its effective coupling to photons is loop-induced by charged fermions and is further controlled by the small component of $H_1$ along the SM-like direction $\Phi_3$. In our benchmark, this overlap is quantified by $R_{31}\equiv\langle \Phi_3|H_1\rangle$. The next-to-lightest state $H_2$ is aligned with the SM Higgs direction and accounts for the observed \SI{125}{GeV} signal.

From a broader perspective, most recent interpretations of the $\gamma\gamma$ feature around $m_{\gamma\gamma}\simeq 95~\mathrm{GeV}$ embed the putative resonance in minimal extensions of the Two-Higgs-Doublet Model. A common strategy is to supplement the 2HDM with an additional gauge-singlet (pseudo) scalar (sometimes with a second singlet), or to introduce an inert sector, so that the neutral spectrum contains mass eigenstates that are admixtures of singlet and doublet fields. In many such constructions a (softly-broken) $\mathbb{Z}_2$ symmetry is imposed to enforce natural flavour conservation and suppress tree-level FCNCs in the Yukawa sector, while in other scenarios this role is replaced or complemented by a gauged $U(1)^\prime$ symmetry (e.g.\ $U(1)_H$ or flavour-dependent $U(1)_X$), which further restricts the scalar potential and Yukawa structure and may require extra chiral matter for anomaly cancellation~\cite{Aguilar-Saavedra:2023tql,Ge:2024rdr,Baek:2024cco,Benbrik:2025wkz,Krishnan:2025haa,Hmissou:2025uep,Arcadi:2025grl,Dong:2025exu}. Complementary studies show that diphoton anomalies---including the \SI{95}{GeV} structure---can also be accommodated within 2HDM frameworks without additional singlet scalars, for instance in aligned and phenomenologically constrained setups where the light state remains predominantly doublet-like~\cite{Coutinho:2024zyp,Khanna:2025cwq,Bhatnagar:2025jhh}.

In the present work, the enlarged scalar sector is not introduced ad hoc to fit the excess: it is dictated by the Peccei--Quinn charge assignments required by the flavored-axion construction and by the texture structure of the quark Yukawa matrices. In this setting, the suppression of tree-level FCNCs follows from the same PQ-driven Yukawa pattern together with the hierarchy $v_{1,2,4}\ll v_3$, rather than from imposing a $\mathbb{Z}_2$ symmetry. The complex singlets $S_{1,2}$ primarily implement PQ breaking and (as discussed in the next section) generate the Majorana masses of right-handed neutrinos, while the collider phenomenology of the \SI{95}{GeV} state is governed dominantly by the doublet sector and its controlled admixture with $\Phi_3$.

\section{Axion–Neutrino Mass Relation\label{sec:axion-neutrino}}
In our model, after spontaneous symmetry breaking, the Yukawa interaction that generates the neutrino mass is given by
\begin{equation}
-\mathcal{L}_{Y^\nu}=y_{ij}^\nu \frac{v_{SM}}{\sqrt{2}}\bar{\nu}_{Li} \nu_{Rj}+\frac{1}{2}\, Y^{N}_{ij}\, \overline{\nu^{\,c}}_{\!\!Ri}\, \nu_{Rj}\frac{v_{s_2}}{\sqrt{2}} + \text{h.c.}
\end{equation}
induces Majorana masses for the right-handed neutrinos once the scalar field \( S_2 \) acquires a vacuum expectation value (VEV), \( v_{s_2} \). Consequently, the right-handed neutrino mass matrix is given by \(m_{ij}^{N}=\frac{1}{\sqrt{2}}Y^{N}_{ij} v_{s_2} \),. In a similar way the Dirac neutrino mass matrix is $m_{ij}^{D}=Y_{ij}^{\nu}v_{\text{SM}}/\sqrt{2}$, where $v_{\Phi_3}\sim v_{\text{SM}}\sim 246$GeV, in such a way that the left-handed neutrinos mass matrix is 
\begin{align}
m^{\nu}_{ij}=-\left(m_D m_N^{-1}m_D\right)_{ij}
=-\frac{v_{\text{SM}}^2}{\sqrt{2}v_{s_2}}\left(Y^{\nu} Y_N^{-1}Y^{\nu}\right)_{ij}
\end{align}

  At the same time, as discussed in the previous section, the axion decay constant is given by \( f_a \simeq v_{s_2} x_{s_2} \).
 From \eqref{eq:mafa} $m_a=(7.55\times 10^{-2}\text{GeV})^2/f_a$ $\simeq \frac{(7.55\times 10^{-2}\text{GeV})^2}{x_{s_2}v_{s_2}} $, in such a way that the neutrino mass matrix is proportional to the axion mass, i.e., 
 $m^{\nu}_{ij}=-x_{s_2}\left(Y^{\nu} Y_N^{-1}Y^{\nu}\right)_{ij}\times 7.5\times 10^{6} m_{a}$.
 in the standard model  $(Y^{\nu})^2$ ranges from $\sim 10^{-11}$ for the electron to $1$ for the top quark, which implies a wide range for the axion masses.  
The  neutrino  mass matrix must be less than the cosmological bound (i.e., $\sum_i m_{\nu_i}\leq$ 0.12eV), which implies 
 $\lvert x_{s_2}\text{Tr}\left(Y^{\nu} Y_N^{-1}Y^{\nu}\right)\times 10^{6} m_{a}\rvert\le 0.12$eV. 
On the other hand, the trace of the neutrino mass determinant is the product of the three neutrino masses, i.e., $m_1^{\nu}m_2^{\nu}m_3^{\nu}=-\frac{\left(\text{det}Y^{\nu}\right)^2}{\text{det}Y^{N}} \left(x_{s_2} 7.5\times 10^{6} m_{a}\right)^3 $.
Combining the  neutrino oscillation data and the cosmological upper bound on the neutrino masses  it is posible to bound 
the neutrino mass determinant $\text{det}(m^{\nu})=m^{\nu}_1m^{\nu}_2m^{\nu}_3$, 
\begin{align}
0\le \lvert m^{\nu}_1m^{\nu}_2m^{\nu}_3 \rvert\le 
\begin{cases}
(0.038\text{eV})^{3},\hspace{0.4cm}\text{N.O.}\\
(0.034\text{eV})^{3},\hspace{0.4cm}\text{I.O.}
\end{cases},
\end{align}
which implies 
\begin{align}
\left|x_{s_2}\frac{\left(\text{det}Y^{\nu}\right)^{2/3}}{\left(\text{det}Y^{N}\right)^{1/3}} \right|m_a
\le 
\begin{cases}
5.1\times10^{-9}\text{eV},\hspace{0.4cm}\text{N.O.}\\
4.5\times10^{-9}\text{eV},\hspace{0.4cm}\text{I.O.}
\end{cases}
\end{align}

\section{Low energy constraints\label{sec:constraints}}
\subsection{Flavor changing neutral currents}
Since our setup involves non-universal PQ charges, a tree-level assessment of flavor-changing neutral currents (FCNCs) is required in addition to the standard bounds on the axion–photon coupling. As emphasized in Ref.~\cite{DiLuzio:2020wdo}, the most stringent limits on axion–quark flavor-violating interactions arise from meson decays into invisible final states. At present, the process $K^{\pm}\rightarrow \pi^{\pm}a$ provides the leading constraints on the axion mass, as reported by the NA62 Collaboration~\cite{Fiorenza_NA62_LaThuile_2026} and summarized in Ref.~\cite{DiLuzio:2020wdo}. For the channels $K^{\pm}\rightarrow \pi^{\pm}a$ and $B\rightarrow K^{*}a$, the tree-level FCNC contributions originate from the  fermion kinetic Lagrangian~\cite{Giraldo:2020hwl}.

In our framework, these contributions are assumed to be absent at the Lagrangian level; hence, the resulting FCNC interactions arise exclusively from field redefinitions and scale proportionally with the PQ charges. As shown in Ref.~\cite{Giraldo:2020hwl}, the decay widths of the pseudoscalar mesons $K^{\pm}(B)$ into an axion and either a charged pion ($K^*$) read~\cite{Giraldo:2020hwl}

\begin{align}
 \Gamma(K^\pm\rightarrow \pi^\pm a) =& \frac{m_K^3}{16\pi}\left(1-\frac{m_\pi^2}{m_K^2}\right)^2\lambda_{K\pi a}^{1/2} f^2_0(m_a^2)
 \lvert g_{ads}^V \rvert^2,\notag\\
 \Gamma(B\rightarrow K^{*}a) =& \frac{m_B^3}{16\pi}
 \lambda_{BK^{*}a}^{3/2} A_{0}^2(m^2_a)\lvert g_{asb}^A\rvert^2,  
  \end{align}
where $\lambda_{Mm a}= \left(1-\frac{(m_a+m)^2}{M^2}\right)\left(1-\frac{(m_a-m)^2}{M^2}\right)$
and 
\begin{align}\label{eq:av-couplings}
g_{ad_id_j}^{V,A}= 
\frac{1}{2f_a c^{\text{eff}}_3}\Delta^{Dij}_{V,A}, 
\end{align}
where:
\begin{equation}
 \Delta^{Dij}_{V,A}= \Delta^{Dij}_{RR}(d)\pm \Delta^{Dij}_{LL}(q),
\end{equation}

\noindent with 
\begin{align}
    \Delta^{Fij}_{LL}(q)&= \left(U^D_{L}x_{q}~U_L^{D\dagger}\right)^{ij},\\
    \Delta^{Fij}_{RR}(d)&= \left(U^D_{R}x_{d}~U_R^{D\dagger}\right)^{ij}_.
\end{align}

In the Eq.~\eqref{eq:av-couplings} we normalize the charges with $c^{\text{eff}}_3$ as it is explained in~\cite{Giraldo:2020hwl}. For $m_a \ll 1$~MeV, the form factor satisfies $f_0(m_a^2)\approx 1$~\cite{Carrasco:2016kpy} in $K^{\pm}\rightarrow\pi^{\pm}a$ decays. From Ref.~\cite{Ball:2004ye}, one finds $f_0(m_a^2)\approx 0.33$ for $B^{\pm}\rightarrow K^{\pm}a$, $f_0(m_a^2)\approx 0.258$ for $B^{\pm}\rightarrow\pi^{\pm}a$, and $A_0(m_a^2)\approx 0.374$ for $B^{\pm}\rightarrow K^{*\pm}a$.

\begin{table}
 \begingroup
\renewcommand*{\arraystretch}{1.2} 
 \begin{equation}
 \begin{array}{|l|l|}
 \hline 
 \hspace{0.5cm} \text{Collaboration} & \text{upper bound} \\
 \hline
 \text{N62 Collaboration\cite{Fiorenza_NA62_LaThuile_2026}} 
            &\mathcal{B}\left(K^+\rightarrow\pi^+      a\right)<(9.6.0^{+1.9}_{-1.8})\times 10^{-11}\\
\text{CLEO~\cite{Ammar:2001gi}}                 
            &\mathcal{B}\left(B^\pm\rightarrow \pi^\pm a\right)<4.9\times 10^{-5}\\
 \text{CLEO~\cite{Ammar:2001gi}}
            &\mathcal{B}\left(B^\pm\rightarrow K^\pm   a\right)<4.9\times 10^{-5} \\
 \text{BELLE~\cite{Lutz:2013ftz} }
            &\mathcal{B}\left(B^\pm\rightarrow \rho^{\pm} a\right)<21.3\times 10^{-5} \\                        
 \text{BELLE~\cite{Lutz:2013ftz} }
            &\mathcal{B}\left(B^\pm\rightarrow K^{*\pm}   a\right)<4.0\times 10^{-5} \\          
\hline
 \end{array}
\end{equation}
  \caption{These inequalities come from the window for new physics in the branching ratio uncertainty of the meson decay in a pair $\bar{\nu}\nu$.}
 \endgroup  
 \label{tab:fcn4c}
\end{table}

\begin{figure}[h!]
\begin{center}
\centering 
\begin{tabular}{cc}
 \includegraphics[scale=0.3 ]{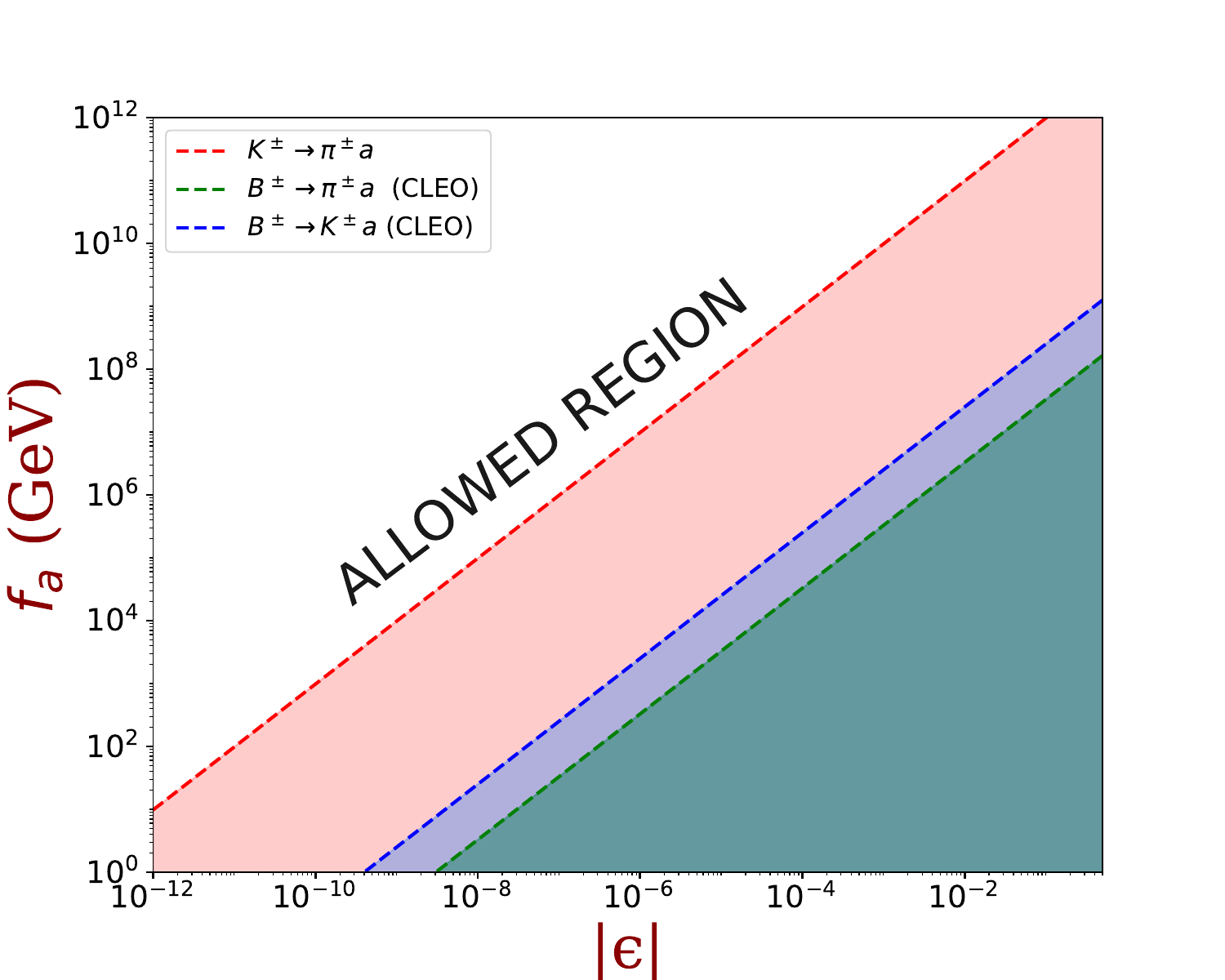}  
\end{tabular}
\end{center}
\caption{Allowed regions by lepton  decays.  For the down-type quarks and charged leptons the non-universal  part of the PQ charges just depend on the diference $s_2-s_1 = N\epsilon/9$, hence the flavor-changing neutral-current couplings (the off diagonal elements)  just depend on $\epsilon$.}
\label{fig:fa_vs_epsilon}	
\end{figure}

The limits on the axion couplings and on the decay constant $f_a$ may be extracted from rare semileptonic transitions $M\rightarrow m \bar{\nu}\nu$, with $M=K^{\pm},B^{\pm}$ and $m=\pi^{\pm},$ $K^{\pm},K^{*\pm},$ and $\rho$. These bounds are collected in Table~\ref{tab:fcn4c}. Figure~\ref{fig:fa_vs_epsilon} displays the dependence of $f_a$ on $\epsilon$. For our charge assignment, the contributions to $B^\pm\rightarrow \pi^{\pm}a$ and $B^\pm\rightarrow K^{\pm}a$ are highly suppressed, leaving $K^{\pm}\rightarrow \pi^{\pm}a$ as the dominant and most constraining channel.

Astrophysical considerations—including black-hole superradiance and the SN~1987A bounds on the neutron electric dipole moment—further restrict the axion decay constant to the range~\cite{DiLuzio:2020wdo} (see Fig.~\ref{fig:fa_vs_epsilon}):$0.8\times10^{6}\text{GeV}\leq f_a\leq 2.8\times 10^{17}\text{GeV}$.

\subsection{Constraints on the axion-photon coupling}
\begin{figure}[h]
\begin{center}
\centering 
\includegraphics[scale=0.35]{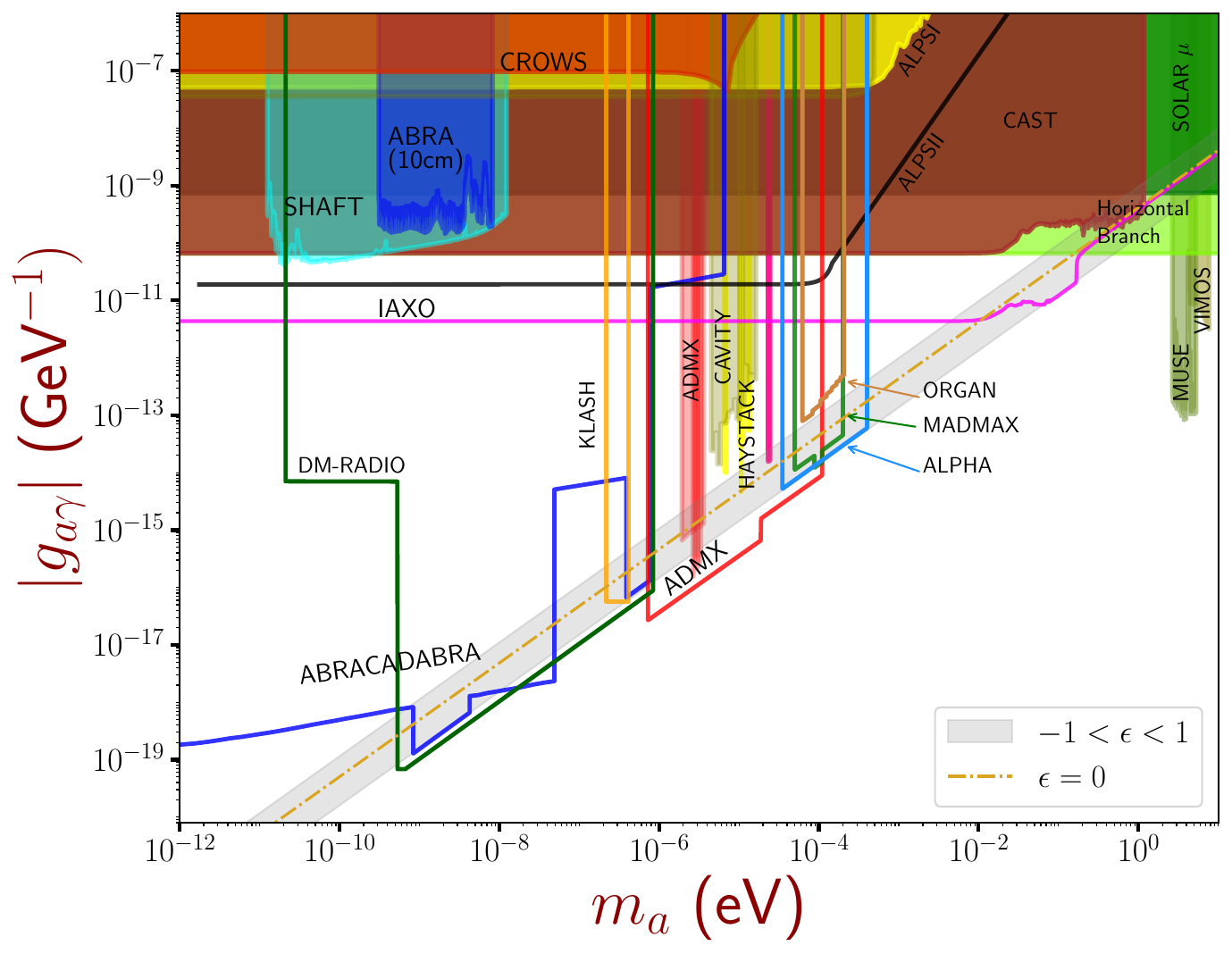}  
\end{center}
\caption{
The excluded parameter space by various experiments corresponds to the colored regions,
the dashed-lines correspond to the projected bounds of coming experiments looking for axion signals \cite{AxionLimits}. The gray region corresponds to the parameter space scanned by our model.
}
\label{fig:ExperimentosAxion}	
\end{figure}
As shown in Appendix~\ref{sec:appendix-axion-photon-coupling}, the low-energy effective Lagrangian contains the axion–photon interaction
\begin{align}
\mathcal{L} \supset -\frac{1}{4} g_{a\gamma} a F_{\mu\nu}\tilde F^{\mu\nu},
\end{align}
which provides one of the most sensitive experimental probes of axions and axion-like particles (ALPs). The corresponding constraints in the $(m_a,g_{a\gamma})$ plane are summarized in Fig.~\ref{fig:ExperimentosAxion}. Below we briefly review the main classes of searches relevant to our parameter space.

Axion searches based on the photon coupling can be grouped into three categories: (i) haloscopes, which target axion dark matter in the Galactic halo; (ii) helioscopes, which search for solar axions; and (iii) purely laboratory experiments based on photon–axion conversion.

\paragraph{Haloscopes.}
Haloscope experiments exploit resonant photon–axion conversion in microwave cavities or broadband magnetized detectors. Leading constraints in the $\mu{\rm eV}$ range are provided by ADMX \cite{ADMX:2021nhd,ADMX:2019uok,ADMX:2020hay}, which has reached sensitivity to QCD-axion models in the mass window around $m_a \sim \mathcal{O}(1\text{--}10)\mu{\rm eV}$. Complementary exclusions at nearby masses have been reported by HAYSTAC \cite{HAYSTAC:2020kwv,HAYSTAC:2018rwy} and by the CAPP collaboration \cite{Lee:2020cfj,Jeong:2020cwz,CAPP:2020utb}. At higher masses ($m_a\sim 100\mu{\rm eV}$), first results from MADMAX further constrain $g_{a\gamma}$, while at ultralow masses ($m_a\lesssim {\rm neV}$) broadband searches such as ABRACADABRA set leading limits. Together, these experiments exclude significant portions of parameter space under the assumption that axions constitute the local dark matter density.

\paragraph{Helioscopes.}
Helioscope searches rely on axion production in the Sun via the Primakoff process and subsequent conversion into x rays in a laboratory magnetic field. The strongest bound to date is provided by CAST,
\begin{align}
g_{a\gamma} < 5.8\times10^{-11}{\rm GeV}^{-1} \quad (95\%~{\rm C.L.}),
\end{align}
for $m_a \lesssim 0.02{\rm eV}$ \cite{CAST:2017uph,CAST:2007jps}. The next-generation experiment IAXO is expected to improve this sensitivity by more than one order of magnitude, probing a substantial fraction of the QCD-axion band. These constraints are independent of the cosmological axion abundance.

\paragraph{Laboratory searches.}
Laboratory-based experiments test photon–axion conversion under controlled conditions, without astrophysical assumptions. Light-shining-through-a-wall experiments such as ALPS-II \cite{Bahre:2013ywa,ALPS:2020} and OSQAR \cite{OSQAR:2015qdv} constrain $g_{a\gamma}$ at low masses, while polarization experiments such as PVLAS \cite{Zavattini:2005tm,Zavattini:2007ee} probe vacuum birefringence and dichroism induced by ALPs. Although currently less sensitive than haloscopes or helioscopes in the QCD-axion region, these searches provide model-independent bounds.

\paragraph{Astrophysical constraints.}
Additional limits arise from stellar energy-loss arguments. Observations of horizontal-branch stars constrain $g_{a\gamma}$ at the level of
$g_{a\gamma}\lesssim 6\times10^{-11}$ ${\rm GeV}^{-1}$ \cite{Ayala:2014pea}, while solar neutrino data and other indirect probes provide complementary bounds \cite{Gondolo:2008dd,Regis:2020fhw,Grin:2006aw}. These constraints are particularly relevant in the sub-eV mass range.

\paragraph{Summary of excluded regions.}
The combined exclusions from haloscope, helioscope, laboratory, and astrophysical observations are displayed in Fig.~3. The figure shows that present data probe a broad range of masses, from $m_a\sim10^{-12}{\rm eV}$ up to $m_a\sim10^{-1}{\rm eV}$, excluding sizeable regions of the $(m_a,g_{a\gamma})$ plane. As discussed below, a nontrivial fraction of our model parameter space overlaps with these experimentally excluded domains.

\section{Discussion and conclusions\label{sec:conclusions}}
In this work, we have presented a realization of a flavored Peccei–Quinn symmetry that simultaneously addresses the strong CP problem, the fermion flavor structure, and the origin of neutrino masses within an extended scalar sector. The framework is constructed by enlarging the scalar content of the Standard Model to include four Higgs doublets and two scalar singlets, supplemented by right-handed neutrinos and a vector-like quark. The imposed PQ symmetry governs both the axion dynamics and the Yukawa structure of the model.

We have shown that the PQ charge assignments allow the generation of Hermitian quark mass matrices with five texture zeros, reproducing the observed quark masses, CKM mixing parameters, and CP-violating phase. 

Neutrino masses arise through a Type-I seesaw mechanism, where the spontaneous breaking of the PQ symmetry generates Majorana masses for right-handed neutrinos. This mechanism leads to an intrinsic relation between the neutrino mass scale and the axion decay constant, providing a direct link between neutrino physics and axion phenomenology.

The scalar sector exhibits a rich mass spectrum containing multiple CP-even, CP-odd, and charged scalar states. The hierarchical pattern of vacuum expectation values naturally aligns one scalar state with the observed 125 GeV Higgs boson while allowing the presence of additional light and heavy scalars. In particular, the model can accommodate a light CP-even scalar near 95 GeV, potentially connected with diphoton excesses reported in collider searches, while heavier scalar states remain consistent with current LHC bounds and provide promising targets for future experimental exploration.

We have analyzed low-energy and astrophysical constraints arising from flavor-changing neutral currents, rare meson decays, and axion-photon interactions. The results indicate that viable regions of parameter space exist where all current experimental limits are satisfied. In this context, semileptonic kaon decays provide the most stringent flavor constraints, while astrophysical observations and axion search experiments impose complementary bounds on the axion decay constant and couplings.

An additional appealing feature of the model is the improved naturalness of the Yukawa sector. Most Yukawa couplings in the quark sector are naturally of order unity, while the neutrino sector requires significantly less fine-tuning compared with the Standard Model expectations. This suggests that flavored PQ symmetries may provide a robust mechanism for explaining fermion mass hierarchies.

Overall, our results demonstrate that flavored PQ constructions with multi-Higgs realizations constitute a viable and predictive framework capable of linking axion physics, flavor structures, and neutrino mass generation. Future collider measurements, flavor experiments, and dedicated axion searches will play a crucial role in probing the parameter space of this class of models and testing the proposed connections between these sectors.

\section*{Acknowledgments} 
We thank Professor Roberto Martínez for introducing us to axion physics. This research was partly supported by the ``Vicerrectoría de Investigaciones e Interacción Social VIIS de la Universidad de Nariño'',  project numbers  3130 and 3595.

\appendix

 \section{Axion-photon Coupling}
 \label{sec:appendix-axion-photon-coupling}
 At energies well below the Peccei--Quinn (PQ) breaking scale, the axion couplings to gauge fields can be derived by performing suitable chiral field redefinitions~\cite{Giraldo:2023osw}. Such transformations leave the classical Lagrangian invariant but induce non-trivial Jacobian factors in the functional measure of the path integral. These anomalous contributions are governed by the divergence of the axial current and generate effective CP-odd interactions between the axion and the gauge-field topological densities. The resulting terms can be written as
\begin{align}
\mathcal{L} &\supset
- c_1^{\text{eff}}\frac{\alpha_1}{8\pi}
\frac{a}{\Lambda_{\text{PQ}}}
B_{\mu\nu}\tilde B^{\mu\nu}
- c_2^{\text{eff}}\frac{\alpha_2}{8\pi}
\frac{a}{\Lambda_{\text{PQ}}}
W_{\mu\nu}^3\tilde W^{3\mu\nu}
\notag\\
&\quad
- c_3^{\text{eff}}\frac{\alpha_3}{8\pi}
\frac{a}{f_a}
G_{\mu\nu}^a\tilde G^{a\mu\nu}
\notag\\
&=
e^2 C_{\gamma\gamma}
\frac{a}{\Lambda_{\text{PQ}}}
F_{\mu\nu}\tilde F^{\mu\nu}
+
\frac{e^2 C_{ZZ}}{c_W^2 s_W^2}
\frac{a}{\Lambda_{\text{PQ}}}
Z_{\mu\nu}\tilde Z^{\mu\nu}
\notag\\
&\quad
+
\frac{2 e^2 C_{\gamma Z}}{c_W s_W}
\frac{a}{\Lambda_{\text{PQ}}}
F_{\mu\nu}\tilde Z^{\mu\nu}
+
g_s^2 C_{GG}
\frac{a}{\Lambda_{\text{PQ}}}
G_{\mu\nu}^a\tilde G^{a\mu\nu}.
\end{align}
In the Standard Model, the neutral electroweak gauge fields are related to the physical photon and $Z$ boson through
\begin{align*}
B^{\mu} &= \cos\theta_W\, A^{\mu} - \sin\theta_W\, Z^{\mu}, \\
W^{3\mu} &= \sin\theta_W\, A^{\mu} + \cos\theta_W\, Z^{\mu},
\end{align*}
where $A^{\mu}$ and $Z^{\mu}$ denote the photon and $Z$-boson fields, respectively. The anomaly coefficients are given by~\cite{Giraldo:2020hwl}
\begin{align}
c_1^{\text{eff}} &= -\frac{1}{3}\Sigma q
+\frac{8}{3}\Sigma u
+\frac{2}{3}\Sigma d
-\Sigma l
+2\Sigma e, \\
c_2^{\text{eff}} &= -3\Sigma q-\Sigma l, \\
c_3^{\text{eff}} &= -2\Sigma q+\Sigma u+\Sigma d-A_Q,
\end{align}
where $\Sigma f \equiv f_1+f_2+f_3$ represents the sum of the PQ charges over the three fermion families.

For phenomenological applications, it is convenient to introduce
\begin{align}
C_{\gamma\gamma}
&= -\frac{1}{32\pi^2}
\left(c_1^{\text{eff}}+c_2^{\text{eff}}\right),\hspace{0.2cm} C_{GG}
= -\frac{1}{32\pi^2}
c_3^{\text{eff}}.
\notag\\
C_{ZZ}
&= -\frac{1}{32\pi^2}
\left(s_W^4 c_1^{\text{eff}}+c_W^4 c_2^{\text{eff}}\right),
\notag\\
C_{\gamma Z}
&= -\frac{1}{32\pi^2}
\left(c_W^2 c_2^{\text{eff}}-s_W^2 c_1^{\text{eff}}\right) . 
\end{align}
It is customary to define the effective PQ scale as
$\Lambda_{\text{PQ}} = |c_3^{\text{eff}}|\, f_a$,
as well as the ratio
\begin{align}
\frac{E}{N}
=
\frac{c_1^{\text{eff}}+c_2^{\text{eff}}}
{c_3^{\text{eff}}}.
\end{align}

The axion--photon coupling then takes the form~\cite{DiLuzio:2020wdo}
\begin{align}
g_{a\gamma\gamma}
=
\frac{4 e^2 C_{\gamma\gamma}^{\text{eff}}}{\Lambda_{\text{PQ}}}
=
-\frac{\alpha}{2\pi f_a}
\left(
\frac{E}{N}-2.03
\right),
\end{align}
with $\alpha = e^2/(4\pi)$. 

Finally, nonperturbative QCD effects associated with the gluon--axion interaction generate a potential for the axion field, which at low energies can be described in terms of axion--pion mixing~\cite{diCortona:2015ldu}. The corresponding axion mass is approximately
\begin{align}\label{eq:mafa}
m_a = 5.7(7)\,\mu\text{eV}
\left(
\frac{10^{12}\,\text{GeV}}{f_a}
\right).
\end{align}
\section{Axion decay constant \texorpdfstring{$f_a$}{} in multi-Higgs models\label{sec:fa}}
Consider $n$ complex scalar fields $\Phi_i$ (indexed by $i=1,\dots,n$) which transform under a global $U(1)_{\rm PQ}$ as
\begin{equation}\label{eq:PQtransf}
\Phi_i\;\to\; e^{i x_i\alpha}\,\Phi_i,
\end{equation}
where $x_i$ are the PQ charges of $\Phi_i$ (integers or rational numbers) and $\alpha$ is a constant and uniform parameter. Each field develops a vacuum expectation value (VEV):
\begin{equation}
\langle\Phi_i\rangle=\frac{v_i}{\sqrt{2}}>0.
\end{equation}
Parametrize fluctuations about the vacuum in polar form (radial + angular decomposition):
\begin{equation}\label{eq:polar}
\Phi_i(x)=\frac{1}{\sqrt{2}}\big(v_i+\rho_i(x)\big)\,e^{i A_i(x)/v_i},
\end{equation}
where $\rho_i(x)$ are the radial (massive) modes and $A_i(x)$ are the angular fields (phases).
The kinetic part of the scalar Lagrangian contains
\begin{equation}\label{eq:Akin}
\mathcal{L}_{\rm kin}=\sum_i |\partial_\mu\Phi_i|^2
\supset \frac{1}{2}\sum_i (\partial_\mu A_i)(\partial^\mu A_i)+\cdots
\end{equation}
where we omitted radial mode kinetic terms and higher order interactions.
To introduce the axion field $a$ we redefined the fields in such a way that 
\begin{align}
 \Phi_i\rightarrow e^{\frac{ax_i}{f_a}}\Phi_i    
\end{align}
That is equivalent to redefining 
\begin{equation}
A_i(x)\rightarrow A_i+\frac{x_i v_i}{f_a}\,a(x)+\ldots
\end{equation}
Inserting this into the kinetic terms~\eqref{eq:Akin} yields
\begin{align}
\mathcal{L}_{\rm kin}\supset& \frac{1}{2}\sum_i\left(\frac{x_i v_i}{f_a}\right)^2(\partial_\mu a)(\partial^\mu a)+\cdots\notag\\
=&\frac{1}{2}\left(\frac{\sum_i x_i^2 v_i^2}{f_a^2}\right)(\partial_\mu a)(\partial^\mu a)+\cdots.
\end{align}
Requiring the canonical normalization for a real field  $\tfrac{1}{2}(\partial_\mu a)^2$ fixes
\begin{equation}\label{eq:fa}
\boxed{\;f_a^2=\sum_{i=1}^n x_i^2 v_i^2\;}
\end{equation}
which matches the well-known result~\cite{Srednicki:1985xd}.


\bibliographystyle{ieeetr}
\bibliography{biblio3,BiblioNoInspire}

\end{document}